# Wide bandgap semiconductors for radiation detection: A review

Ivana Capan

*Ruđer Bošković Institute, Bijenička 54; 10 000 Zagreb, Croatia; capan@irb.hr*

**Abstract:** In this paper, an overview of the wide bandgap (WBG) semiconductors for radiation detection applications is presented. The recent advancements in the fabrication of high-quality wafers have enabled the remarkable WBG semiconductor device applications. The most common 4H-SiC, GaN and β-Ga2O3 devices used for radiation detection are described. The 4H-SiC and GaN devices have already achieved exceptional results in the detection of alpha particles and neutrons, thermal neutrons in particular. While β-Ga2O3 devices have not yet reached the same level of technological maturity (compared to 4H-SiC and GaN), their current achievements for X-ray detection indicate great potential and promising prospects for future applications.

## 1. Introduction

A need, for reliable and efficient radiation detectors for particle physics, space technologies, nuclear power plants, medicine, and homeland security applications, is growing rapidly. The requirements set for radiation detectors are complex, from exceptional efficiency and energy resolution to extreme radiation tolerance. Among numerous candidates, semiconductor radiation detectors offer plenty of advantages due to their exceptional material properties. For many decades, Si-based radiation detectors have been the champions in the radiation detection arena [1]. However, Si-based devices are reaching the limit of their performance, and it is dubious that significant improvements will follow in years to come. Due to the wider bandgap (compared to Si 1.12 eV, for example), and the recent astonishing progress in material fabrication, wide bandgap (WBG) semiconductors are becoming a new driving force for radiation detection. Material properties of WBG semiconductors (Table 1) that are determining for radiation detection applications and make WBG semiconductors suitable for high-temperature and radiation-harsh environments are, among others, wide bandgap, high breakdown voltage, high electron mobility, and exceptional thermal properties [2].

The list of the most scrutinized WBG semiconductors includes but is not limited to silicon carbide (SiC), gallium nitride (GaN), gallium arsenide (GaAs), cadmium telluride (CdTe), and gallium oxide ($Ga_2O_3$). In this review paper, the attention will primarily be focused on the selected materials, namely, SiC, GaN, and $Ga_2O_3$. More precisely, 4H-SiC and β-$Ga_2O_3$. The reasoning is the following, the 4H polytype is the preferred material among the best-known SiC polytypes (2C-, 3C-, 4H- and 6H-SiC) for electronic components due to the high and isotropic mobility of charge carriers [2,3]. Monoclinic β-$Ga_2O_3$ is the most stable among the five crystalline phases of $Ga_2O_3$ single crystals (α, β, γ, δ and ε phases) and poses a very high breakdown electric field (Table 1) [4]. The basic material properties for 4H-SiC, GaN, and β-$Ga_2O_3$ are given in Table 1.

**Table 1.** Material properties for the selected wide bandgap (WBG) semiconductors covered in this review paper (GaN, 4H-SiC and $Ga_2O_3$) [4,5].

|  | GaN | 4H-SiC | β-Ga$_2$O$_3$ |
|---|---|---|---|
| Relative dielectric constant ε | 9.0 | 9.7 | 10.2-12.4 |
| Bandgap energy (*eV*) | 3.4 | 3.3 | 4.5 |
| Density (*g/cm$^3$*) | 6.15 | 3.21 | 6.44 |
| *e-h* pair creation energy (*eV*) | 8.9 | 7.8 | 15.6 [6] |
| Breakdown electric field (*MV/cm*) | 3.3 | 2.5 | 7-8 |
| Electron mobility (*cm$^2$/Vs*) | 1200 | 1000 | ~200 |
| Saturation electron velocity (x10$^7$ cm/s) | 2.5 | 2.0 | 1.0-1.5 |
| Thermal conductivity (*W/cmK*) | 2.1 | 2.7 | 0.11-0.27 |

The main aim of this review paper is to provide an easy-to-follow yet practical and above all useful overview of the recent achievements of WBG semiconductors used for radiation detection. The paper is explicitly focused on radiation detection applications. Other very important aspects such as crystal growth, material characterization, or different applications will not be covered. However, the appropriate references that could be useful for additional reading and better understanding of the subject, will be provided whenever relevant.

The structure of this paper is the following. Three main sections (Sections 2, 3, and 4) are devoted to the selected WBG semiconductors. Namely, Section 2 to 4H-SiC, Section 3 to GaN, and Section 4 to β-Ga$_2$O$_3$. Each section has two main subsections. The first one will provide a short introduction and the second one will provide an overview of the recent advancements, covering radiation response to different radiation sources such as alpha particles, neutrons and X-rays. Section 5 provides a summary of the main challenges and future perspectives.

## 2. 4H-SiC

Research on SiC dates back to the end of the 19th century when SiC was recognized as a material for an abrasive powder and refractory bricks [7]. In the 1950's the SiC potential was again recognized, this time for high-temperature electronic devices [7]. Despite certain efforts through the decades that followed, research on SiC began to flourish by the end of the 20th century. The progress in the fabrication of high-quality 4H-SiC wafers has enabled remarkable 4H-SiC-based device applications: power electronics [8,9], quantum sensing [10,11,12] and radiation detection [13,14,15,16,17,18]. This has influenced the significant increase in the market value. The SiC device market, valued at around $2 billion in 2023, is projected to reach $11 billion to $14 billion in 2030 [19].

There is a whole variety of 4H-SiC-based devices that are currently being used as radiation detectors. The most common are PiN diodes [20], metal-oxide-semiconductor (MOS) structures [21,22], and Schottky barrier diodes (SBDs) [15,16,17]. Even though the SBD is one of the simplest devices, it has many advantages and it has been chosen as the preferred structure in many studies

[23]. Figure 1 shows a scheme of a typical n-type 4H-SiC SBD. In lots of reported studies, Ni is a preferred material for Schottky and Ohmic contacts for the n-type SBDs. However, it should be noted that other metals are also being used. Osvald et al. [24] have recently reported on Ni/Au Schottky contacts and Chen et al. [25] on the possible benefits of Mo Schottky contacts. Lees et al. [26] have made additional changes and used semi-transparent Cr/Ni Schottky contact. As we shall explain later in the text, semi-transparent Schottky contacts could improve the efficiency for the detection of low-energy X-rays, compared to conventional Schottky contacts. More detailed information about the key parameters of 4H-SiC SBDs such as the epitaxial layer thickness, and Schottky contact area could be found elsewhere [23]. In addition, a list of excellent review papers dedicated to SiC has recently been published. De Napoli [27] and Coutinho et al. [3] have provided extensive overviews of crystal growth, material properties, and characterization techniques, with a dedicated focus on SiC radiation detection applications.

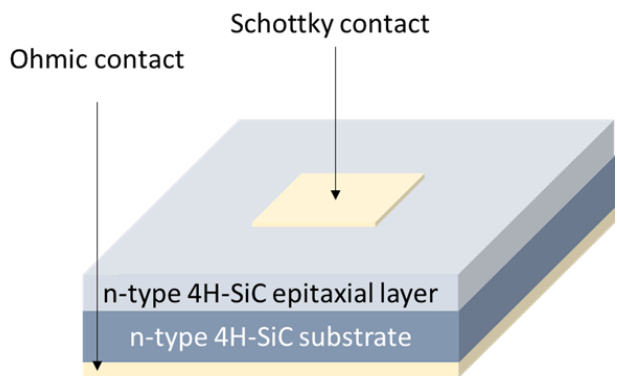

**Figure 1.** A scheme of the typical n-type 4H-SiC SBD used for radiation detection.

*2.1. Radiation response to alpha particles and neutrons*

Almost all research on the radiation response starts with laboratory tests using alpha particles from $^{241}$Am source. 4H-SiC-based devices are not an exception. The early and yet still significant work in the area of 4H-SiC radiation response to alpha particles was done by Ruddy et al. [14]. Using the various alpha emitters in the 3.18–8.38 MeV energy range, an excellent energy resolution was achieved. Through the years, progress in energy resolution and efficiency has been reported by many authors [28,29,30,31]. Chaudhuri et al. [31] achieved an excellent energy resolution of 0.29% FWHM for 5.48 MeV alpha particles. Additionally, the same group has applied 4H-SiC-based MOS structures and compared them to 4H-SiC SBDs. They have reported the highest energy resolution ever measured on SiC-based MOS detectors: 0.42% for 5.48 MeV alpha particles [32]. While most of the radiation tests are done at room temperature (RT), Bernat et al. [33] have recently investigated the 4H-SiC SBDs radiation response to alpha particles at elevated temperatures in a vacuum. They have used n-type 4H-SiC SBDs with an active area of 25mm². Figure 2. shows the response to alpha particles ($^{241}$Am source with a characteristic alpha particle maximum of 5.48 MeV) measured at different temperatures (200 – 390 K) in a cryostat under a vacuum (<0.1 mbar) [33]. As seen in Figure 2, the peak maximum is shifted on the x-scale as the radiation temperature increases. The estimated energy resolution was 2.5% and did not change with the temperature.

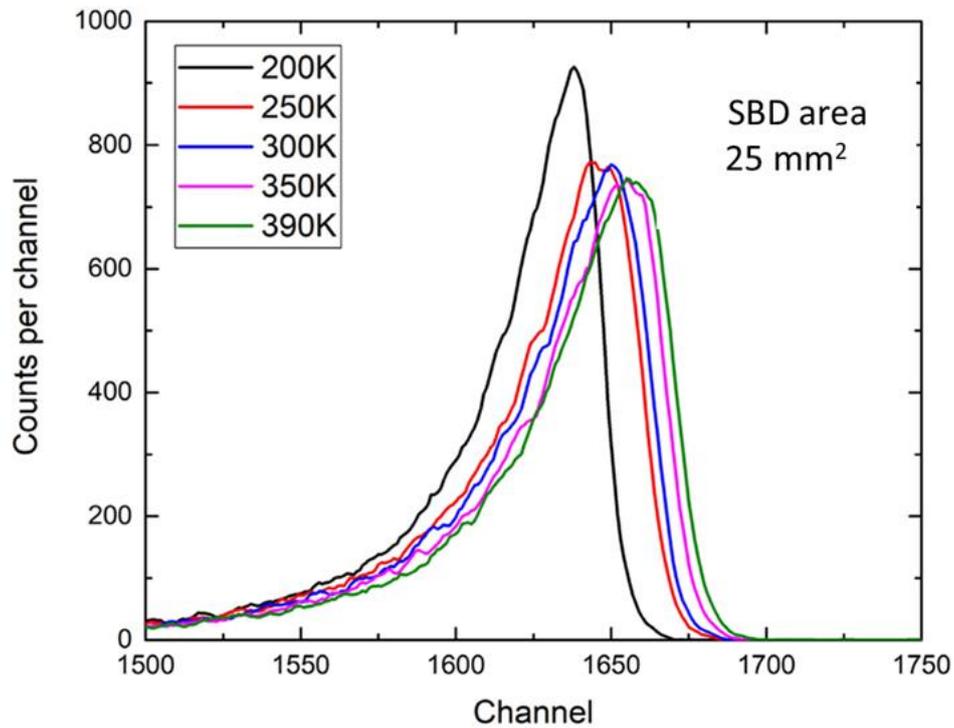

**Figure 2.** Radiation response of 4H-SiC SBDs (active area of 25 mm$^2$) to alpha particles ($^{241}$Am source) in a vacuum at different temperatures. Data reproduced from Ref. [33].

As already mentioned before, due to the specific material properties WBG semiconductors are suitable for radiation harsh environments, such as the International Thermonuclear Experimental Reactor (ITER) [34], for example. The significant advantage of SiC lies in the fact that SiC can detect and distinguish both, thermal and fast neutrons. Detection of thermal and fast neutrons by SiC-based devices differs, as thermal neutrons could not be directly detected. Thermal neutron presence is obtained from the detection of ionizing neutron reaction products, such as alpha particles and tritons. On the contrary, fast neutrons could be directly detected due to the elastic scattering of fast neutrons with Si or C atoms, or indirectly using polyethylene-based converters. Possible neutron-induced reactions with Si and C that could participate in the 4H-SiC detector response are $^{12}$C(n,n)$^{12}$C and $^{28}$Si(n,n)$^{28}$Si [35]. The probability of this scattering increases as the detector's active layer thickness increases. The prospect of detecting 14 MeV fast neutrons by 4H-SiC detectors was demonstrated by F.H. Ruddy et al. [36]. Another work is also worth mentioning, Flamming et al. [18] used 2.5 MeV neutrons and measured the radiation response with and without polyethylene converters. Better results are achieved using the polyethylene converters.

Hitherto, 4H-SiC SBDs have mostly been used for thermal neutron detection. As already said, they cannot be directly measured, therefore effective thermal neutron converters are needed. The requirement for such converters is that they are rich in isotopes with a large cross-section for neutrons with energy in the range of kBT at RT. The frequently used converters are $^6$Li and $^{10}$B [37]. Figure 3 shows a typical set-up for thermal neutron detection using the 4H-SiC SBD with the thermal neutron converter placed just above the Schottky contact (a few mm above). The converter has been horizontally shifted in Figure 3, for clarity.

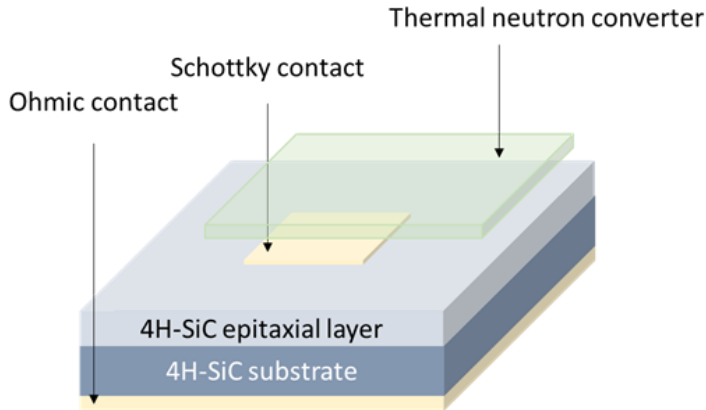

**Figure 3**. A scheme of 4H-SiC SBD with the additional thermal neutron converter placed above the Schottky contact. The converter is used for thermal neutron detection and is usually placed a few mm above the Schottky contact.

The best-reported efficiencies for thermal neutrons detection using the 4H-SiC devices is between 4-5% [38,39,40]. Recently, Bernat et al. [33] have reported an efficiency of 5.02% with the use of a 26.54 μm thick $^6$LiF thermal neutron converter layer (Figure 4). Additionally, Bernat et al. [33] have compared the different SBD active area sizes and concluded that with the increase in the SBD active area, the detector could register thermal neutrons with a nuclear reactor power as low as 1kW.

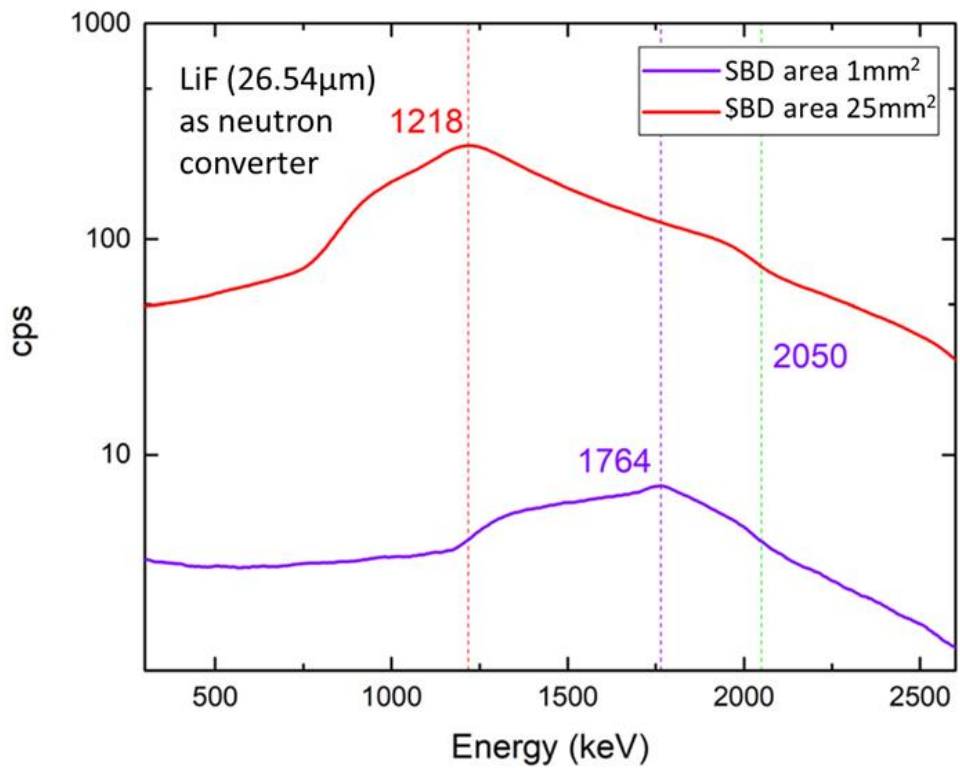

**Figure 4.** Radiation response of two 4H-SiC SBDs with different active areas (1 mm² and 25 mm²). The 26.54 μm thick $^6$LiF thermal neutron converter layer was placed above the SBDs, as already described in the text. Data were taken from Ref. [33].

Contrary to the alpha particles and neutrons, the low energy X-rays and γ-rays detection by 4H-SiC devices has not yet reached the same level of efficiency. However, several attempts have been made, and they should be noted. Mandal et al. [41] have achieved a full width at half maximum (FWHM) of 1.2 keV at 59.6 keV using the n-type 4H-SiC SBDs. Moreover, using the same SBDs, they were able to detect low-energy X-rays in the energy range of 13.93 - 26.20 eV. Lees et al. [26] have used a slightly different SBD structure. They have reduced the thickness of the Schottky contact, from a typical 50-100 nm down to 18 nm, and prepared so-called semi-transparent SBD. With such a 4H-SiC device, an FWHM of 1.47 eV at 22 keV is achieved.

## 3. GaN

Like its WBG counterpart SiC, GaN is a well-known material as the research on GaN dates to the 19th century. The gallium and its compounds were discovered by Paul-Émile Lecoq de Boisbaudran in 1875 [42]. The GaN research flourished in 1969 when H.P. Maruska and J.J. Tietjen reported on the growth of single-crystal film of GaN [43], and again in 1972, when J.I. Pankove, E.A. Miller, and J.E. Berkeyheiser developed a GaN-based blue light detector [44]. The number of GaN-based devices for optoelectronics, power electronics, and radio frequency (RF) applications is constantly growing [45] which leads to a GaN device market value increase. Radiation detection applications are still far less exploited, compared to power electronics. Figure 5 shows a few GaN devices used for radiation detection, such as (a) SBD, (b) double SBD and (c) metal-semiconductor-metal (MSM). As it is for n-type 4H-SiC, the most common Schottky contacts are Au and Ni. More detailed information about GaN material properties and radiation detection could be found in a comprehensive review paper by Wang et al. [5] and references therein.

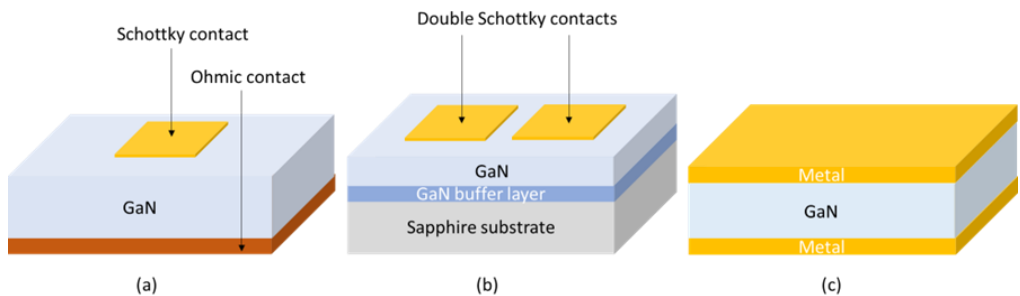

**Figure 5**. Schemes of the typical GaN devices used for the detection of alpha particles and neutrons, (a) SBD, (b) double SBD, and (c) MSM. Figure adapted from Ref. [5].

*3.1. Radiation response to alpha particles and neutrons*

The pioneering work on alpha particle detection was done by Vaikuts et al. [46]. They have achieved charge collection efficiency (CCE) of 92% for 5.84 MeV alpha particles ($^{241}$Am source) using the double SBD as shown in Fig. 5b. Additionally, Muligan et al. [47] have fabricated GaN SBDs (as shown in Fig. 5a) for measuring the response for alpha particles ($^{241}$Am source). They have obtained excellent results and a charge collection efficiency of 100 %.

Figure 6 shows the radiation response to alpha particles obtained by GaN SBD at different voltages (from − 400 up to −750 V) in a vacuum. CCE of 100% for 5.48 MeV alpha particles was achieved at −750 V [48].

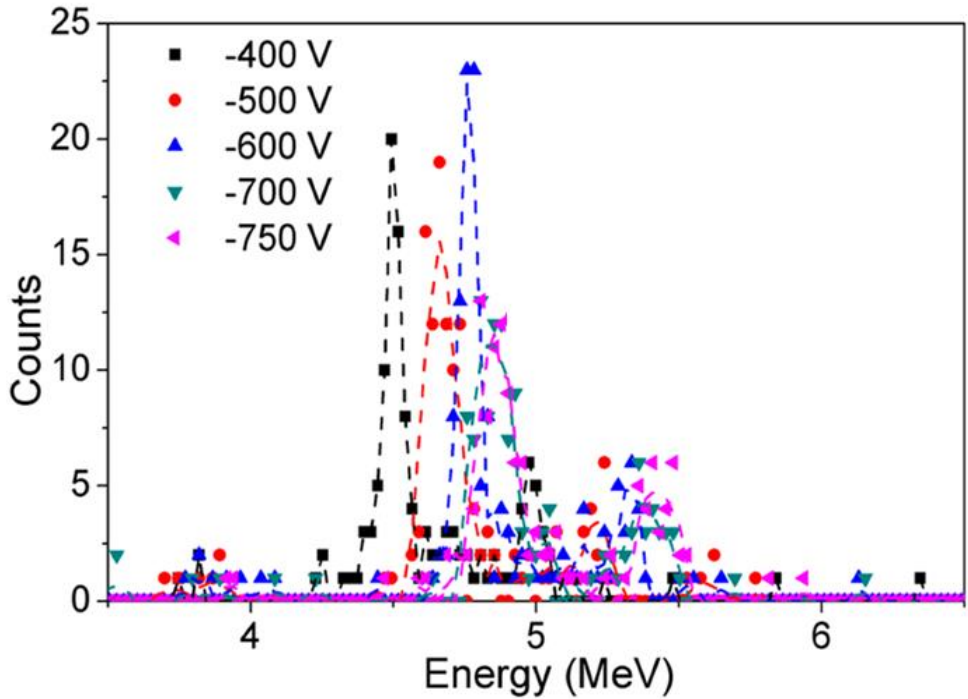

**Figure 6**. Radiation response of GaN SBD to alpha particles ($^{241}$Am source) at different voltages in vacuum. Data taken from Ref. [48].

While all the above-mentioned results are obtained at RT, Zhu et al. [50], have performed the temperature-dependent (290 – 450 K) radiation response measurements to alpha particles using the GaN PiN structure. They have observed that the peak maximum is shifted on the x-scale (i.e. energy-scale) as the temperature increases, which is almost identical to what Bernat et al. [33] have observed with 4H-SiC SBDs.

Another similarity between 4H-SiC and GaN devices is their ability to detect neutrons. Detection could be direct and indirect by using converters made of $^6$Li, $^{10}$B and $^{157}$Gd [50]. In the case when neutron converters are used, the main principle is identical to what has been previously described for 4H-SiC SBD (Figure 3). The converter is placed just above the Schottky contact.

An interesting prospect that GaN SBD could be used for direct neutron detection through the reaction $^{14}$N(n,p)$^{14}$C, which could make the neutron converters redundant has recently been presented by Zhou et al. [50]. Figure 7 shows the gamma and thermal neutron response of Si-doped GaN scintillators exposed to the reactor gamma rays and thermal neutrons. The neutron-induced peak (red spectrum) has been attributed to ionization from 584 keV protons produced by the $^{14}$N(n, p)$^{14}$C reaction.

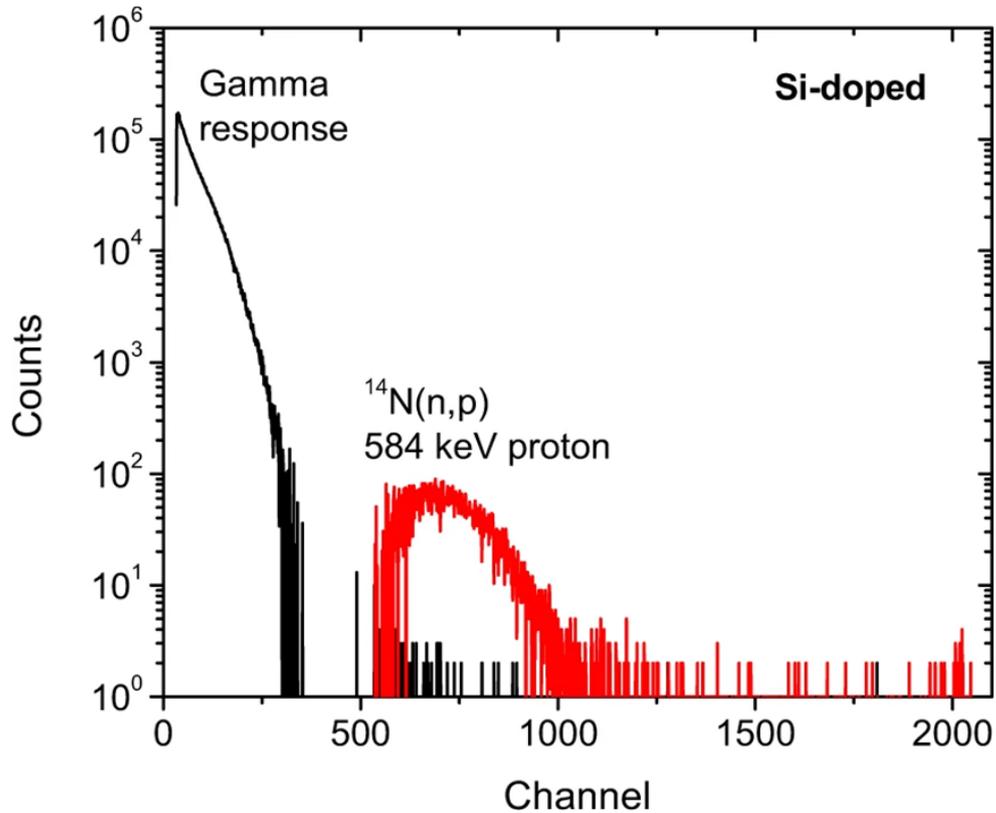

**Figure 7.** Radiation response to gamma and thermal neutrons measured by Si-doped GaN scintillator without thermal neutron converter. Data taken from Ref. [50].

The number of published research papers related to the GaN X-ray detectors is noticeably lower compared to the GaN alpha particles or neutron detectors. Here, we mention results obtained by Duboz et al. [51]. They have used GaN MSM devices with Pt/Au metallic contacts to test radiation response to X-rays. The absorption coefficient in GaN MSM devices was measured as a function of the photon energy in the X-ray range from 6 to 40 keV. They have concluded that GaN could be used as an X-ray detector, but only for energies below 20 kV.

## 4. β-Ga2O3

Research on $Ga_2O_3$ shares a similar history as on GaN. Upon the discovery of gallium in 1875, significant improvements were achieved in the 1960's. The band gap of bulk single crystals of $Ga_2O_3$ was estimated as 4.7 eV. In the following decades, the major challenge and limiting factor was the quality of $Ga_2O_3$ crystals [52]. Due to the low crystal quality, the $Ga_2O_3$-based applications were not developed at the same pace as SiC and GaN-based devices.

As previously mentioned, due to the stability and high breakdown electric field β-$Ga_2O_3$ is the preferred among $Ga_2O_3$ phases and is used for different applications. The number of applications is continuously growing over the past ten years. β-$Ga_2O_3$ has shown potential for power electronics, solar-blind ultraviolet (UV) photodetectors, and gas sensors [52]. Moreover, like all WBG semiconductors, β-$Ga_2O_3$ has significant potential for applications in harsh environments (high radiation, high temperature, high voltage). Today, the high-power market is dominated by SiC devices, but β-$Ga_2O_3$ is the most promising candidate to take over the ultra-high-power market [53].

Nevertheless, β-$Ga_2O_3$-based devices for radiation detection have not yet reached the same level of maturity as 4H-SiC and GaN devices. However, early, and very encouraging results on radiation response have been obtained for β-$Ga_2O_3$ X-ray detection. Developing a highly efficient, sensitive, and reliable X-ray detector for medical imaging and homeland security is still a very challenging task. An excellent overview of the recent advancements in β-$Ga_2O_3$-based X-ray detectors and scintillators has recently been published by Prasad et al. [54]. A comparison of the X-ray detector's crucial parameters such as X-ray-generated photocurrent, response time, response, sensitivity and signal-to-noise ratio is presented in detail.

Figure 8. shows the preferred β-Ga2O3 devices used for radiation detection, (a) MSM and (b) SBD. Different metals have been used for Schottky contacts like tungsten (W), copper (Cu), nickel (Ni), iridium (Ir), platinum (Pt) and gold (Au) [55]. Pt and Ni are the most used among them.

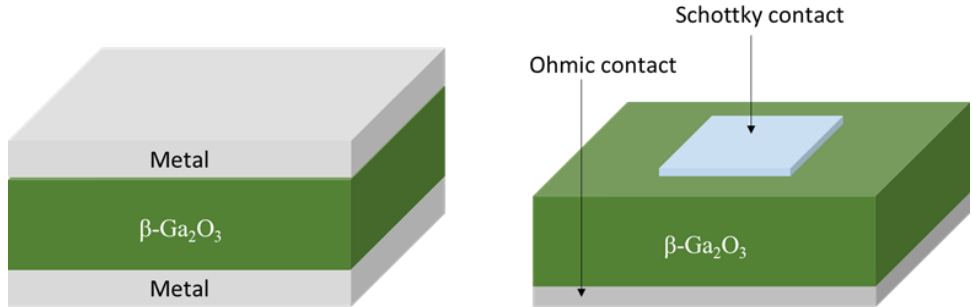

**Figure 8**. β-Ga2O3 devices for radiation application, (a) MSM and (b) SBD.

*4.1. Radiation response to X-rays*

Compared to 4H-SiC and GaN, the number of research studies on the radiation response of β-Ga2O3 devices to alpha particles and/or neutrons is rather limited. However, it is rather impressive to follow the advancements of β-Ga2O3 devices used for X-ray detection. Here, we will give an overview of the recent results. One of the most used devices is the Fe-doped β-Ga2O3 MSM. Hany et al. [56] have obtained promising results (response time 0.3s) and proposed that doping β-Ga2O3 with Fe could significantly improve the detector performance. Additional improvements were made by Chen et al. [57]. They have also fabricated X-ray detectors on Fe-doped β-Ga2O3 MSM. Again, Ti/Au metal electrodes were deposited on both sides of the Fe-doped Ga2O3 samples. The detector showed great potential for X-ray detection with a slightly shorter response time (0.2s) and a high sensitivity.

Lu et al. [58] have investigated the β-Ga2O3 SBDs with Pt/Au as Schottky contacts for X-ray detection. The X-ray source with a peak photon energy at around 24 keV was used for measuring radiation response. Figure 9 shows the radiation response to different incident fluxes (different flux was achieved by adjusting the X-ray tube current, as shown in Figure 9). This result has again indicated the great potential of β-Ga2O3 devices for X-ray detection.

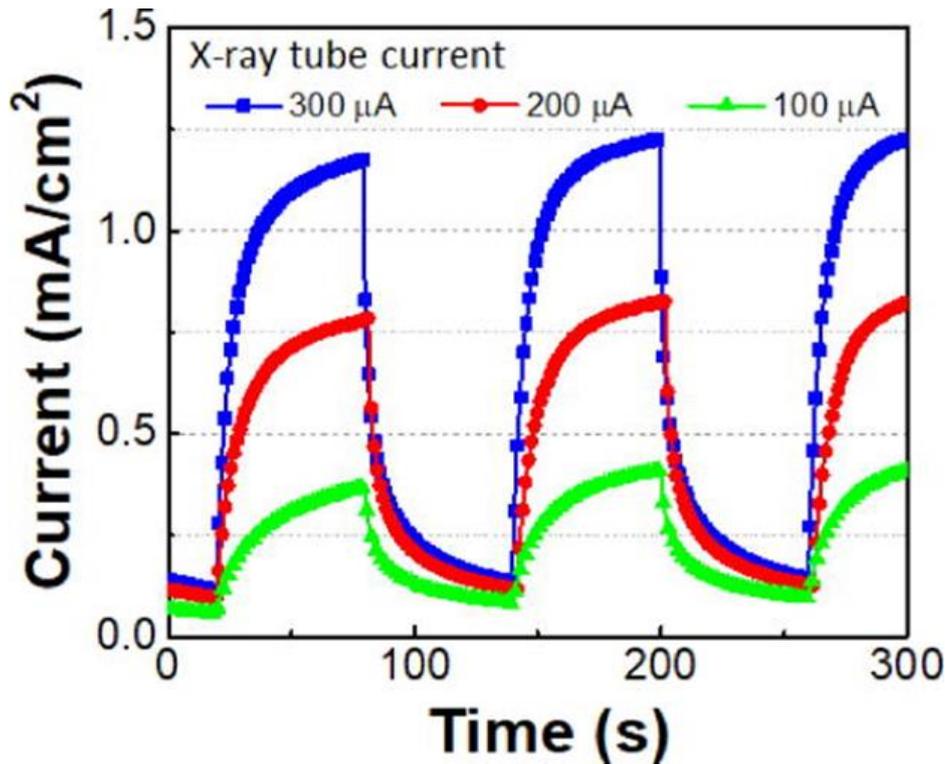

**Figure 9.** Radiation response of the β-Ga2O3 SBD to X-ray with different incident flux (controlled by the X-ray tube current). Reprinted from Appl. Phys. Lett. **112**, 103502 (2018) with the permission of AIP Publishing.

At the end, we will briefly mention a different but still promising approach. Zhang et al. [59] have used the Sn-doped β-Ga$_2$O$_3$ microwires and fabricated X-ray detectors. The fabricated detector exhibits stability over extended temperature ranges, from room temperature to 623 K which is one of the highest reported operating temperatures for β-Ga$_2$O$_3$ X-ray detectors.

## 5. Conclusions

The main aim of this paper is to present recent advancements and challenges in the application of WBG semiconductors for radiation detection.

Despite the evident advancements, radiation detection applications of the WBG semiconductors did not reach the same level of development as in the case of power electronics, where 4H-SiC and GaN are the main driving forces. Moreover, the fabrication of high-quality β-Ga$_2$O$_3$ wafers has yet to overcome several difficulties (for example defect-free material). Once the crystal quality has improved, the performance of β-Ga2O3 devices will follow, as was previously the case with 4H-SiC.

It is clear, that the most mature technology is related to 4H-SiC and GaN devices, where exceptional results for the detection of alpha particles and neutrons, thermal neutrons in particular, are achieved. Different 4H-SiC and GaN devices are used, but the most relevant results are obtained with a very simple device such as SBD. While 4H-SiC and GaN devices have not yet reached the desired efficiency for the detection of X-rays, that makes a possible niche for β-Ga2O3 devices to prosper. The available results on β-Ga2O3 devices for X-ray detection strongly support this.

Despite all the challenges, the future perspectives for WBG semiconductors are bright. For radiation detection applications, 4H-SiC, GaN, and β-Ga$_2$O$_3$ form a set of materials that complement each other efficiently and enable the development of detectors that will cover a wide range of radiation (alpha particles, thermal and fast neutrons, X-rays).